# *Nuzzer*: A *Large-Scale* Device-Free Passive Localization System for Wireless Environments

Moustafa Seifeldin
Wireless Intelligent Networks Center (WINC)
Nile University
Cairo, Egypt
Email: moustafa.sefeldin@nileu.edu.eg

Moustafa Youssef
Department of Computer Science
University of Maryland
College Park, MD 20742, USA
Email: moustafa@cs.umd.edu

*Abstract*—The widespread usage of wireless local area networks and mobile devices has fostered the interest in localization systems for wireless environments. The majority of research in the context of wireless-based localization systems has focused on device-based active localization, in which a device is attached to tracked entities. Recently, device-free passive localization (*DfP*) has been proposed where the tracked entity is neither required to carry devices nor participate actively in the localization process. *DfP* systems are based on the fact that RF signals are affected by the presence of people and objects in the environment. The *DfP* concept enables a wide range of applications including intrusion detection and tracking, border protection, and smart buildings automation. Previous studies have focused on small areas with direct line of sight and/or controlled environments. In this paper, we present the design, implementation and analysis of *Nuzzer*, a *large-scale* device-free passive localization system for *real* environments.

*Nuzzer* is designed to satisfy specific goals; high accuracy, ubiquitous coverage, scalability, and operation in real environments. Without any additional hardware, it makes use of the already-installed wireless data networks to monitor and process changes in the received signal strength (RSS) transmitted from access points at one or more monitoring points. We present probabilistic techniques for *DfP* localization and evaluate their performance in a typical office building, rich in multipath, with an area of 1500 square meters. Our results show that the *Nuzzer* system gives device-free location estimates with less than 2 meters median distance error using only two monitoring laptops and three access points. This indicates the suitability of *Nuzzer* to a large number of application domains.

## I. INTRODUCTION

With mobile devices and wireless networking becoming more and more pervasive in our daily lives, context aware applications have gained huge interest. As one of the main context information, location determination has been an active area of research. Therefore, Many localization systems have been proposed, including the GPS system [1], ultrasonic-based systems [2], infrared-based systems (IR) [3], and RF-based systems [4]. All these systems share the requirement of attaching a device to the tracked entity. Recently, we proposed the device-free passive localization (*DfP*) concept [5]. A *DfP* system provides the capability of tracking entities not carrying any devices nor participating actively in the localization process. This is particularly useful in applications such as intrusion detection, border protection, and smart homes automation.

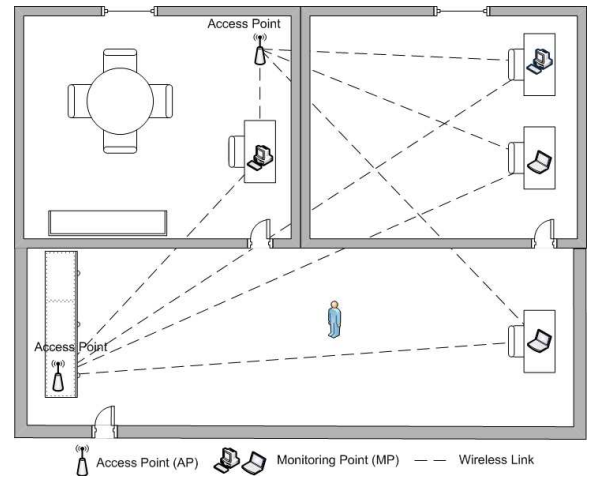

Fig. 1. An example of the different components of a device-free passive localization system in a typical office environment. APs represent signal transmitters. Standard laptops and wireless-enabled desktops represent monitoring points. Any device can be used as an application server.

The *DfP* concept is based on the idea that the existence of an entity, e.g. a human, in an RF environment affects the RF signals, especially when dealing with 2.4 GHz band common in wireless data networks, such as WiFi and WiMax.

A typical *DfP* system consists of (Figure 1): (1) signal transmitters, such as access points (APs) and stations used in typical WiFi deployments, (2) monitoring points (MPs), such as standard laptops and wireless-enabled desktops, along with (3) an Application server (AS) for processing and initiating actions as needed.

A few systems have been introduced for *DfP* localization in wireless environments [5], [6] with a focus on small areas with direct LOS and/or controlled environments [1]. In this paper, we present the design, implementation and analysis of *Nuzzer*, a ***large-scale*** device-free passive localization system for ***real*** environments, rich in multipath (Figure 2).

Although *Nuzzer* can operate in both indoor and outdoor environments, we focus in this paper on the more challenging case of indoor environments. In indoor environments, LOS paths from the transmitters to the receivers are usually ob-structed by walls. In addition, indoor environments contain substantial amounts of metal and other reflective materials that affect the propagation of RF signals in non-trivial ways, causing severe multipath effects. Generally, reflection, refraction, diffraction, and absorption of RF signals result in multipath fading, which may either decrease or increase the RSS at the MPs. Moreover, RF signals are also affected by noise, interference from other sources, and interference between channels. Sources of interference include radio-based transmission devices, microwave ovens, cordless phones, and Bluetooth devices. This makes the problem of indoor localization challenging, especially for the *DfP* case.

---
[1]We discuss related work in more details in Section IV.

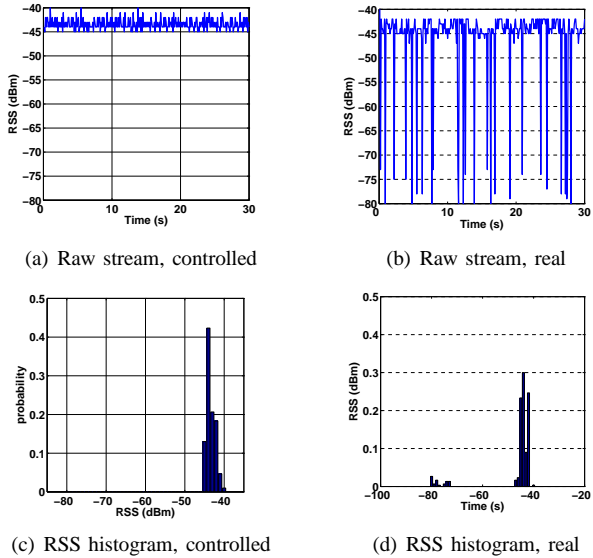

Fig. 2. RSS behavior in a controlled versus a real environment.

The *Nuzzer* system aims at achieving specific goals: high accuracy, ubiquitous coverage, scalability to large areas, and operation in real environments.

### A. Approach

In order to perform localization, we need to capture the relation between signal strength and distance. Since this relation is very complex in indoor environments [7], we do this using a "**passive**" radio map. A radio map is a structure that stores information of the signal strength at different locations in the area of interest [8], [9]. This is usually constructed only one time during an offline phase. Note that passive radio maps differ from active radio maps used in device-based active localization systems. We highlight these differences in Section II.

During the online phase, the *Nuzzer* system uses the signal strength samples received from the APs at the monitoring points and compares them to the passive radio map to estimate the location of the tracked entity.

Radio map based techniques used in device-based active localization can be categorized into two broad categories: deterministic techniques and probabilistic techniques. Deterministic techniques, represent the signal strength of an AP at a certain location by a scalar value, such as the mean value. Then non-probabilistic approaches are used to estimate the location of the tracked entity. For example, in the RADAR system [4] nearest neighborhood techniques are used to infer the user location. On the other hand, probabilistic techniques, e.g. [10], store information about the signal strength distributions from the APs in the radio map. Then probabilistic techniques are used to estimate the location of the tracked entity. Probabilistic techniques for device-based active localization systems are known to give better accuracy [11].

In the *Nuzzer* system, we propose probabilistic techniques to implement *DfP* localization in large-scale real environments and show how they differ from device-based active localization techniques. ***We focus on the problem of localization of a single intruder and leave the general problem of multiple-entities localization to a future paper.***

### B. Contribution

The contribution of this paper is four fold:
1) We present a probabilistic approach for handling the device-free passive localization problem for a single intruder.
2) We present post processing techniques to enhance the accuracy of the basic probabilistic technique.
3) We evaluate the performance of the proposed techniques in a large-scale typical office environment, rich in multipath.
4) We study the effect of changing the systems parameters on the localization process.

### C. Paper Organization

Section II presents the different algorithms used in the *Nuzzer* system and the difference between device-based and device-free localization. Section III presents the evaluation of the *Nuzzer* system in a large-scale typical office environment and the effect of the different parameters on performance. Section IV presents a comparison between *Nuzzer* and the relevant related work. Finally, Section V concludes the paper and gives directions for future work.

## II. THE *Nuzzer* SYSTEM

In this section, we present the different algorithms used in the *Nuzzer* system. We start by an overview of the system followed by a description of our probabilistic algorithms.

### A. Overview

The *Nuzzer* system works in two phases:
- Offline phase: where we build the passive radio map. A passive radio map is similar to the active radio map usually used in device-based active WLAN location determination systems, such as [4], [10], [12]. However, in an active radio map, a user stands with a device at the radio map locations and collects samples from all the APs in range. On the other hand, for the passive radio map construction, a user stands at the radio map locations, without carrying any device, and his effect on the different data streams received at the MPs is recorded. Figure 3 demonstrates the difference between active and passive radio map construction.
- Online Phase: where we estimate the user location based on the RSS from each data stream and the passive radio map prepared in the offline phase.

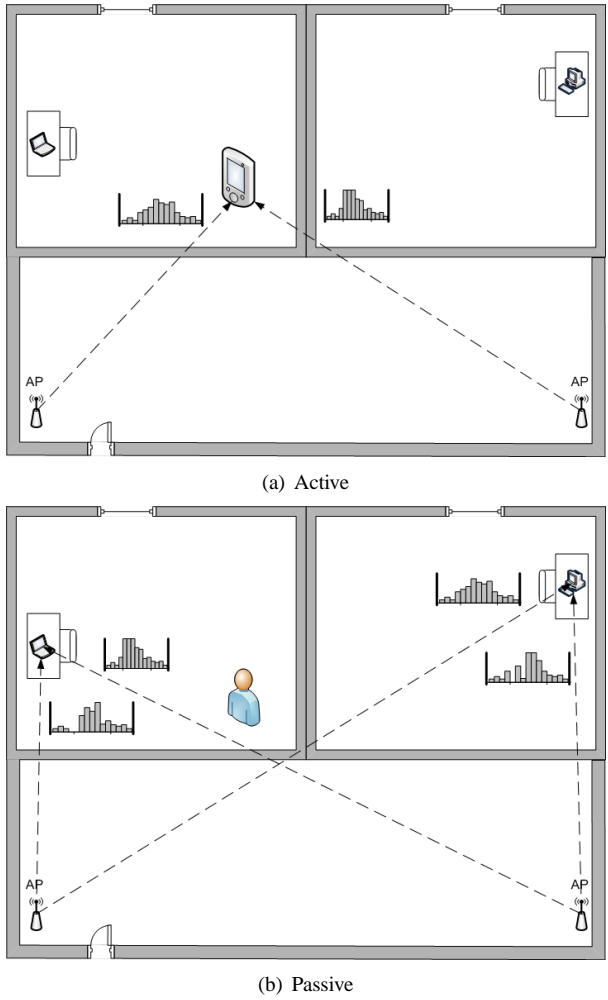

Fig. 3. Difference between active and passive radio maps construction. In a passive radio map, we have a histogram per raw data stream, as compared to a histogram per AP. Also, a user does not carry any device when constructing the passive radio map.

We define two modes of operation for the online phase: The Discrete Space Estimator and the Continuous Space Estimator.
- The Discrete Space Estimator module returns the radio map location that has the maximum probability given the received signal strength vector from different streams. Therefore, the output of the discrete space estimator must be one of the calibrated locations.
- The Continuous Space Estimator works as a post processing step after the discrete space estimator and tries to return a more accurate estimate of the user location in the continuous space. Therefore, if a user is standing between two radio map locations, the continuous space estimator should provide a better estimate than the discrete space estimator.

We start by presenting our mathematical model followed by details of the two modes of operation.

*B. Mathematical Model*

Without loss of generality, let $\mathbb{X}$ be a two dimensional physical space. Let $q$ represent the total number of data streams in the system (number of APs multiplied by number of MPs). We denote the $q$-dimensional signal strength space as $\mathbb{Q}$. Each element in this space is a $q$-dimensional vector whose entries represent the signal strength readings from different streams, where each stream represents an (access point, monitoring point) pair. We refer to this vector as $s$. We also assume that the samples from different APs are independent and hence, the samples of different streams are independent. A user standing at any location $x \in \mathbb{X}$ affects the signal received at the different MPs, and hence the equivalent $q$ dimensional vector.

Therefore, the problem becomes, given a signal strength vector $s = (s_1, ..., s_q)$, we want to find the location $x \in \mathbb{X}$ that maximizes the probability $P(x|s)$.

In the next section, we assume a discrete space $\mathbb{X}$. We discuss the continuous space case in Section II-D.

*C. Discrete Space Estimator*

During the offline phase, *Nuzzer* estimates the signal strength histogram for each stream corresponding to the user standing at each radio map location. Therefore, at each radio map location, we have a set of histograms representing the signal strength received from each stream when the user stands at this location (Figure 3(b)).

Now, consider the online phase. Given a signal strength vector $s = (s_1, ..., s_q)$, one entry per stream, we want to find the location $x \in \mathbb{X}$ that maximizes the probability $P(x|s)$, i.e., we want

$$argmax_x[P(x|s)] \qquad (1)$$

Using Bayes' theorem, this can be written as:

$$argmax_x[P(x|s)] = argmax_x[P(s|x).\frac{P(x)}{P(s)}]$$
$$= argmax_x[P(s|x).P(x)] \qquad (2)$$

Assuming that all locations are equally probable [2], the term $P(x)$ can be factored out from the maximization process in Equation 2. This yields:

$$argmax_x[P(x|s)] = argmax_x[P(s|x)] \qquad (3)$$

$P(s|x)$ can be calculated using the histograms constructed during the offline phase as:

$$P(s|x) = \prod_{i=1}^{q} P(s_i|x) \qquad (4)$$

[2]If the user profile, $P(x)$, is known, i.e. the probability of the user being at each of the radio map locations, it can be used in Equation 2.

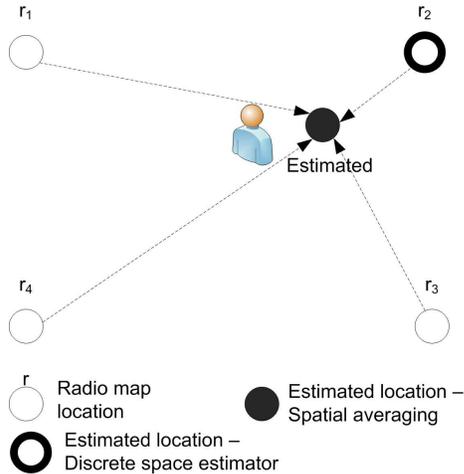

Fig. 4. An example of using the spatial averaging technique to enhance accuracy. The discrete space estimator will return the location $r_2$ as it is the nearest to the actual user location. Using the spatial averaging technique, a better location estimate can be obtained by calculating the center of mass of the top 4 locations ($k = 4$).

The above equation considers only one sample from each stream for a location estimate. In general, a *number of successive samples*, $m$, from each stream can be used to improve performance.

In this case, $P(s|x)$ can then be expressed as follows:

$$P(s|x) = \prod_{i=1}^{q} \prod_{j=1}^{m} P(s_{i,j}|x) \quad (5)$$

Where $s_{i,j}$ represents the $j^{th}$ sample from the $i^{th}$ stream.

Thus, given the signal strength vector $s$, the discrete space estimator applies Equation 5 to calculate $P(s|x)$ for each location $x$ and returns the location that has the maximum probability.

### D. Continuous Space Estimator

The discrete space estimator returns a single location from the set of locations in the passive radio map. In general, an entity need not be standing at one of the radio map locations. Therefore, to increase the system accuracy, *Nuzzer* uses spatial and time averaging techniques to obtain a location estimate in the continuous space.

*1) Spatial averaging:* This technique is based on treating each location in the radio map as an object in the physical space whose weight is equal to the probability assigned by the discrete space estimator, normalized so that the sum of probabilities equals one. We then obtain the center of mass of the $k$ objects with the largest mass, where $k$ is a system parameter, $1 \leq k \leq \parallel \mathbb{X} \parallel$. Figure 4 shows an example of using the spatial averaging technique.

More formally, let $P(x)$ be the probability of a location $x \in \mathbb{X}$, i.e., the radio map, and let $\overline{\mathbb{X}}$ be the list of locations in the radio map *ordered* in a descending order according to the normalized probability assigned from the discrete space estimator. The center of mass technique estimates the current location $x$ as:

$$x = \frac{\sum_{i=1}^{k} P(i).\overline{\mathbb{X}}}{\sum_{i=1}^{k} P(i)} \quad (6)$$

Note that the estimated location $x$ need not be one of the radio map locations.

*2) Time averaging:* This technique uses a time averaging window to smooth the resulting location estimates. The technique obtains the location estimate by averaging the last $w$ location estimates obtained by either the discrete space estimator or the spatial averaging estimator.

More formally, given a stream of location estimates $x_1, x_2, ..., x_t$, the technique estimates the current location $\overline{x_t}$ at time $t$ as:

$$\overline{x_t} = \frac{\sum_{i=t-min(w,t)+1}^{t} x_i}{min(w,t)} \quad (7)$$

The length of the time averaging window affects the latency and accuracy of the system as discussed in Section III.

## III. PERFORMANCE EVALUATION

In this section, we study the performance of the proposed discrete space estimator and continuous space estimator. We start by describing the experimental setup and data collection, followed by studying the effect of different parameters on the performance of the proposed techniques. We also compare the performance of our system to two other estimators:

1) A random estimator: this is used as a baseline for performance comparison. A random estimator selects a random location in the area of interest as its estimate.
2) A deterministic technique: Based on the RADAR system, this estimator stores in the radio map the average signal strength from each stream at each location. During the online phase, the deterministic estimator returns the radio map location whose stored signal strength vector is closest, in signal strength space, to the received vector. More details about this technique can be found in the accompanying technical report [13].

### A. Experimental Testbed

Our experimental testbed is located in the first floor of a two-storey typical office building (Figure 5). The floor has an area of 1500 sq. m. (about 16000 sq. ft.). The experiment was carried out in the main entrance and the corridors, where there were furniture, plants, and substantial amount of metal.

Our experiment was conducted in an 802.11b environment, which operates at 2.4 GHz frequency band. The building had ten Cisco APs (model 1130). For our experiment, we selected only three APs which cover the first floor. We also used two different laptops; one Dell Latitude D830, and one HP Pavilion ze5600 laptop. The two laptops had Orinoco Silver cards attached to them. APs represent the transmitting units, while laptops represent the MPs. Figure 5 shows the locations of APs and MPs.

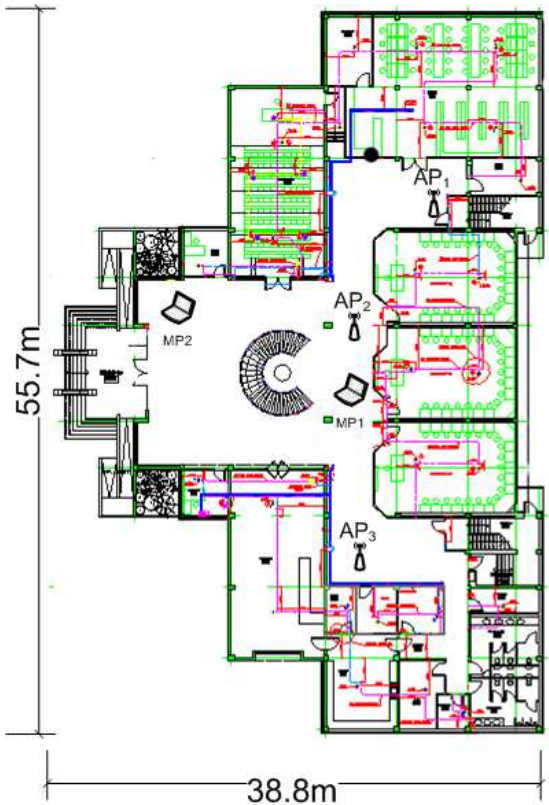

Fig. 5. Floor plan of the area where the *DfP* experiment was performed. The environment is rich in multipath, where furniture, plants, and substantial amount of metal exist. The figure also shows the locations of APs and MPs.

TABLE I
TUNABLE PARAMETERS USED IN OUR EXPERIMENTS

| Parameter | Default value | Meaning |
|---|---|---|
| $n$ | 6 | Number of processed raw data streams |
| $m$ | 26 | Number of consecutive samples to use from one stream per location estimate |
| $k$ | 2 | Number of locations to average in the spatial averaging technique |
| $w$ | 5 | Size of the time averaging window |

| Technique | $25^{th}$ perc. | $50^{th}$ perc. | $75^{th}$ perc. |
|---|---|---|---|
| Probabilistic | 1.2m | 2.9m | 8.98m |
| Deterministic | 3.86m ($3.2\times$) | 8.4m ($2.9\times$) | 13.2m ($1.5\times$) |
| Random | 8.8m ($7.3\times$) | 14m ($4.8\times$) | 18.8m ($2\times$) |

TABLE II
COMPARISON BETWEEN THE $25^{th}$, $50^{th}$, $75^{th}$ PERCENTILE VALUES OF DISTANCE ERROR FOR DIFFERENT *discrete space* ESTIMATOR TECHNIQUES. THE TABLE SUMMARIZES INFORMATION IN FIGURE 6. NUMBERS BETWEEN BRACKETS INDICATE THE DEGRADATION OF DETERMINISTIC AND RANDOM TECHNIQUES COMPARED TO PROBABILISTIC TECHNIQUE.

### B. Data Collection

The wireless cards measure different physical signals during the experiment, such as signal strength and noise. We use only the received signal strength indicator (RSSI) values, reported in units of dBm, which is known to be a better function of distance than noise [4]. We used the active scanning technique, which is part of the 802.11 standard [14], to collect samples from the access points at the rate of five samples per second.

Each one of the two MPs records samples from the three APs, giving a total of six data streams (one stream for each (MP, AP) pair). During the offline phase, a person stands at each of these 53 different locations and we record the samples for 60 seconds for each of the six data streams, giving a total of 300 samples per stream.

For testing purposes (online phase), we collected another *independent* test set at 32 locations. The test set was collected at a different time from the training set. We use this test set to obtain all figures in this section. During the offline and online phases there was no body in the building except the person being tracked. Without loss of generality, we consider a fixed orientation for the person being tracked throughout the experiment.

### C. System Parameters

For the discrete space estimator, we can tune the number of consecutive samples to use from each stream ($m$). Similarly, we can tune the number of raw data streams to use ($n$).

For the continuous space estimator, in addition to these two parameters, we can tune the number of locations to use in the spatial averaging ($k$) and the length of the window to use for time averaging ($w$). Table I summarizes the parameters used in our system. *Unless otherwise specified, we use the default parameters values ($n = 6, m = 26, k = 2, w = 5$), which give the best combined performance.*

### D. Discrete Space Estimator

Figure 6 shows the cumulative distribution function (CDF) of the distance error using the Discrete Space Estimator. We have a total of six data streams, corresponding to the three APs and two MPs we used. Table II summarizes the results of the figure. It lists the $25^{th}$, $50^{th}$, $75^{th}$ percentile values of the distance error. We can see from the figure that the median distance error of the discrete space estimator is 2.9m meters, 2.9 times better than deterministic techniques and 4.8 times better than the random estimator. This ratio is even more for the lower percentile values.

The value of the CDF at zero distance error indicates the probability of determining the exact location.

*1) Impact of the number of samples per stream:* Figure 7 shows the effect of increasing the number of samples used from each stream per location estimate on the accuracy of the system (parameter $m$). The figure shows that, as expected, the median distance error decreases as $m$ increases. However, as $m$ increases, the latency, i.e. time required per location estimate, of the system increases as we have to wait till we collect the $m$ samples. Therefore, a balance is required between the accuracy and latency of the system. This depends on the specific deployment environment. Another approach is to use a moving window of $m$ samples, where at each estimate, one new sample is added to $m - 1$ old samples. This removes the requirement of waiting for $m$ samples.

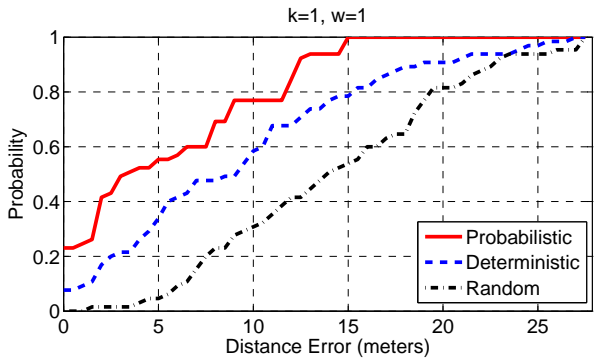

Fig. 6. CDFs of the Euclidean distance between actual locations and locations estimated by the *discrete space* estimator techniques: the probabilistic estimator, the deterministic estimator, and the random estimator.

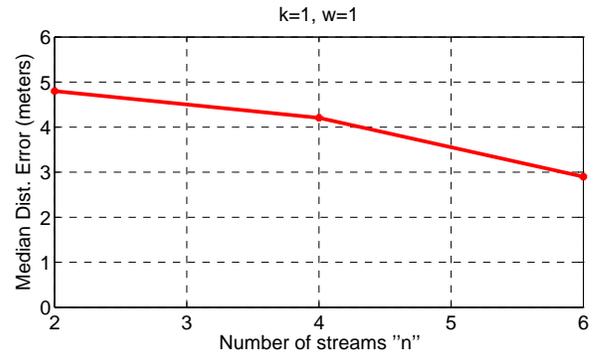

Fig. 8. Median distance error of the discrete space estimates versus the number of used raw data streams $n$. For a given $n$, the figure reports the best median distance error over all the $\binom{6}{n}$ raw streams combinations.

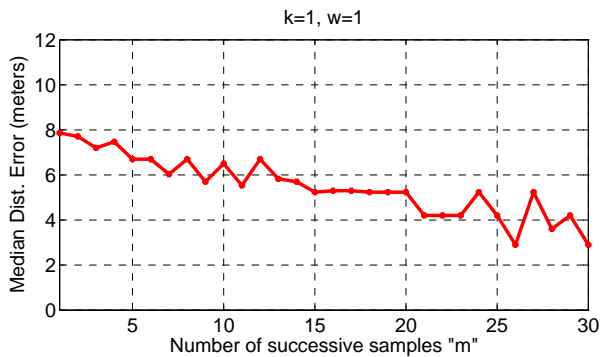

Fig. 7. Median distance error of the discrete space estimator for different values of the number of successive samples from each stream per location estimate ($m$).

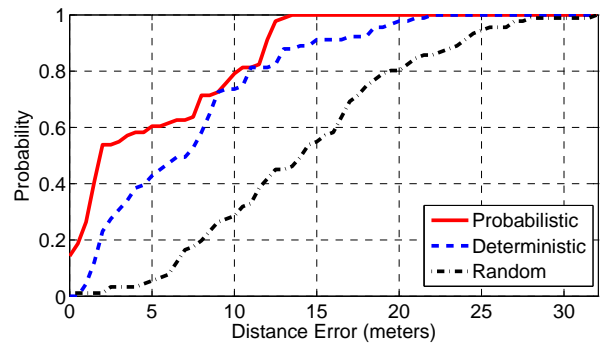

Fig. 9. CDFs of the Euclidean distance between actual locations and locations estimated by *continuous space* estimation techniques: the probabilistic estimator, the deterministic estimator, and the random estimator.

*2) Impact of the number of streams:* Figure 8 shows the median distance error versus the number of streams $n$ used in the estimation process. For a specific $n$, we plot the best result over all possible $\binom{6}{n}$ combinations of streams. The figure shows that as the number of streams increases, we have more information about the environment, and thus we can obtain better accuracy.

### E. Continuous Space Estimator

Figure 9 shows the cumulative distribution function (CDF) of the distance error using the Continuous Space Estimator for the best values of the parameters. Table III summarizes the results of the figure. We can see from the figure that the median distance error of the discrete space estimator is ***1.82*** meters, 3.7 times better than deterministic techniques and 7.7 times better than the random estimator.

*1) Spatial averaging:* We now discuss how the system performance is affected by the number of neighboring locations ($k$) included in spatial averaging technique. Figure 10 shows the effect of increasing the number of neighbors used in the spatial averaging process ($k$) on the median distance error. The figure shows an improvement of 10% between $k = 1$ and $k = 3$. The proposed technique is not sensitive to the increase in $k$, for large $k$, because as we increase $k$ the estimated conditional probability of the locations decreases significantly and hence their effect on the location estimate decreases.

*2) Time averaging:* Figure 11 shows the effect of increasing the size of the time averaging window ($w$) on the median distance error. The figure shows that an improvement of 65% for $w = 5$ as compared to $w = 1$. Again, we have a tradeoff between accuracy and latency. The higher the value of $w$, the higher the accuracy and the higher the latency.

### F. Summary

In this section, we showed that using only six data streams, the *Nuzzer* system provides a *non-LOS DfP* localization system capable of covering large areas, rich in multipath, with very high accuracy; 1.82 meters median distance error. Although this accuracy is lower than the accuracy reported by device-based active localization systems (0.5 meters in [10]), it still suitable for a wide class of applications.

| Technique | $25^{th}$ perc. | $50^{th}$ perc. | $75^{th}$ perc. |
|---|---|---|---|
| *Nuzzer* | 1.2m | 1.82m | 9.5m |
| Deterministic | 2.37m (1.96x) | 6.74m (3.7x) | 10.6m (1.1x) |
| Random | 8.8m (7.3×) | 14m (7.7×) | 18.8m (2×) |

TABLE III
COMPARISON BETWEEN THE $25^{th}$, $50^{th}$, $75^{th}$ PERCENTILE VALUES OF DISTANCE ERROR USING DIFFERENT *continuous space* ESTIMATION TECHNIQUES. THE TABLE SUMMARIZES INFORMATION IN FIGURE 9. NUMBERS BETWEEN BRACKETS INDICATE THE DEGRADATION OF DETERMINISTIC AND RANDOM ESTIMATORS COMPARED TO PROBABILISTIC ESTIMATOR.

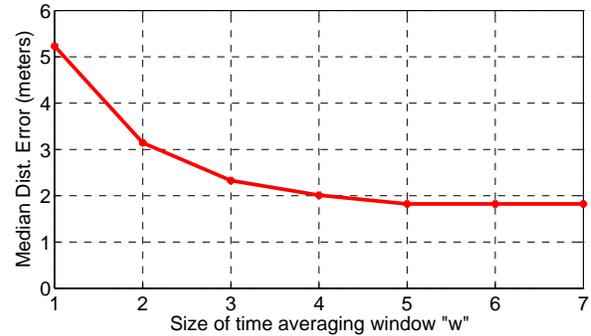

Fig. 11. Median distance error of the continuous space estimates versus the time averaging window size ($w$).

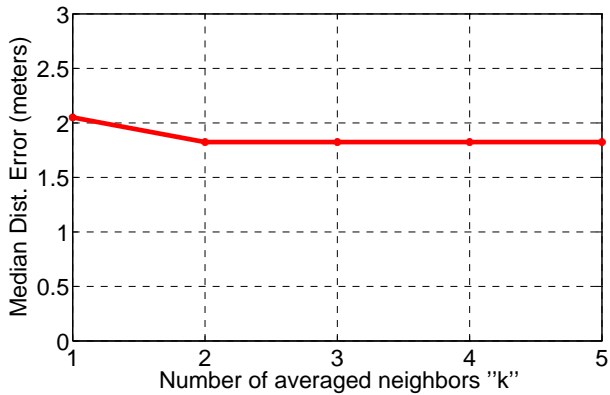

Fig. 10. Median distance error of the continuous space estimator versus the number of neighbors used in the spatial averaging ($k$).

Comparing the performance of the continuous space estimator to the discrete space estimator, we find that the median distance error in the discrete space is 2.9 meters, whereas in the continuous space, the median is 1.82 meters, 37% better.

The spatial averaging and temporal averaging techniques are independent and can be used together to further enhance performance. Combining all techniques, leads to the above mentioned accuracy.

The system parameters $m$ and $w$, which represent the number of samples from each stream and the time averaging window size respectively, can be tuned to balance accuracy and latency, depending on the deployment environment.

The results also showed that the *Nuzzer* system can provide very good accuracy, even when the number of available data streams is as low as two streams. This shows the usability of the system in environments with limited hardware installment, such as in homes.

## IV. RELATED WORK

This section discusses relevant related work. We start by the device-based active localization systems followed by other device-free passive localization systems.

### A. Device-based Active Localization

A number of systems has been introduced over the years to address the localization problem. These systems include the GPS system [1], ultrasonic-based systems [2], infrared-based systems [3], and RF-based systems [4]. All these systems share the requirement that the tracked entity needs to carry a device. In addition many of these technologies require the device being tracked to actively participate in the localization process by running part of the localization algorithm. Moreover, some of these systems are limited in range due to the physical characteristics of the signal they use in localization.

*Nuzzer* allows entities tracking without them carrying any device nor participating actively in the localization process. In addition, *Nuzzer* works with the standard wireless data networks, and thus enhances the value of the data network. Since RF signals penetrates walls, *Nuzzer* does not require LOS and has good coverage range.

### B. Device-free Passive Localization

A number of systems over the years have considered device-free passive localization, including computer vision systems, physical contact systems, radar based systems, and medical imaging based systems.

Using video cameras is a traditional way for passive localization of human beings. For example, [15] describes algorithms for detecting and tracking multiple people in cluttered scenes using multiple synchronized cameras located far away from each other. However, video cameras fail to work in the dark and in presence of smoke. In addition, they suffer from occlusion problems and cannot track entities that are out of sight, limiting their range and scalability.

Physical contact systems, for example the Smart Floor system [16], track the person based on his contact with the environment. For example, in [16], the system uses pressure sensors to detect the presence of a person over floor tiles.

|  | **MIMO Radar-based Systems** | **Radio Tomographic Imaging (RTI)** | *Nuzzer* **System** |
|---|---|---|---|
| Measured Physical Quantity | Reflection and scattering | RSS attenuation | Changes in RSS |
| Range (based on frequency) | Short | Long | Long |
| Wall penetration | Very high | High | High |
| non-LOS localization | Yes | No | Yes |
| Number of deployed nodes (or devices) | Few | Many | Few |
| Complexity of single node (or device) | High | Low | Moderate |
| Number of streams | N/A (echo based) | Large (756) | Low (6) |
| Covering large areas | Limited by its short range (high frequency) | Limited by number of deployed nodes (LOS) | Yes |
| Accuracy | Very High | High | High |
| Accuracy degrades significantly with multipath | No | Yes | No |
| Handles a number of entities | Yes | Yes | Ongoing work |
| Licence-free frequency band | No | Yes | Yes |
| Special hardware required | Yes | Yes | No |

TABLE IV
COMPARISON OF DIFFERENT RF-BASED PASSIVE LOCALIZATION SYSTEMS

These systems require special set up and hardware, and therefore, their scalability is limited.

Ultrawideband (UWB) radar systems provide "Through-wall" detection and tracking. UWB radar systems can utilize impulse [17], frequency-modulated continuous-wave (FMCW) [18], stepped frequency [19], or noise [20] waveforms. These systems are very accurate, yet very complex. An alternate development is to use a Doppler radar with a two-element receiver array to provide less complexity [21]. This Doppler radar assumed that no two targets have the same Doppler return, which is not valid in case of human tracking since micro Doppler returns from the human arm and leg motions have a broad Doppler spread [22]. A four-element array radar can also be used [23]. This latter combines Doppler processing with software beamforming to resolve targets along both the Doppler and direction of arrival (DOA) space.

MIMO radar employs multiple transmit waveforms and have the ability to jointly process the echoes observed at multiple receive antennas ( [24] and references therein). Elements of the MIMO radar transmit independent waveforms resulting in an omnidirectional beampattern. It can also create diverse beampatterns by controlling correlations among transmitted waveforms. In MIMO, different waveforms are utilized and can be chosen to enhance performance in a number of ways.

In summary, radar-based systems are able to provide accurate location estimates. However, they require special hardware and their high complexity limits their applications.

Another emerging technology is Radio Tomographic Imaging (RTI) [6]. It presents a linear model for using RSS measurements to obtain images of moving objects. The proposed system uses *hundreds* of raw data streams obtained from sensor nodes. The system measures the attenuation in the transmitted signal rather than scattering and reflection. Since this system is based on LOS, its accuracy degrades as multipath components increase. To overcome multipath, a higher density of nodes is used.

The concept of *DfP* localization was first introduced in [5]. Experiments were set up in a highly *controlled* and small environment. In addition, the user was allowed to move in only one dimension. Results show that the system can track the intruder's position with more than $86\%$ accuracy in this limited controlled environment. These results have established the proof of feasibility of the *DfP* concept.

The *Nuzzer* system has unique characteristics that differentiate it from the previous systems: It gives high accuracy for large-scale typical environments; it does not require any special hardware; it does not require LOS to operate; and it works with a very low number of raw data streams.

Table IV summarizes the differences between *Nuzzer* and the recent *DfP* RF-based localization systems.

## V. CONCLUSIONS

We presented the design, implementation, and evaluation of the *Nuzzer* device-free passive localization system. *Nuzzer* uses the standard wireless data networks installed in the environment to monitor and process the RSS at one or more monitoring points leading to estimating the location of entities, without requiring them to carry any devices. It works by constructing a passive radio map during an offline phase, then uses a Bayesian-based inference algorithm to estimate the most probable user location given the received signal strength vector and the constructed radio map.

We also presented two post processing techniques: the spatial and temporal averaging to further enhance the accuracy of the basic Bayesian-based algorithm. Using these techniques, the performance of the *Nuzzer* system was enhanced by $38\%$.

We evaluated the performance of the *Nuzzer* system in a typical office building, rich in multipath, with an area of more than 1500 square meters. We used two laptops and three access points. Our results show that the *Nuzzer* system gives a median distance error of $1.82$ meters, 3.7 times better than deterministic techniques and 7.7 times better than a random estimator.

*The presented techniques allow* Nuzzer *to achieve its goals of high accuracy and operation in real environments. By*

*working with the standard wireless equipment,* Nuzzer *also inherits the scalability and ubiquitous coverage of the current wireless technologies.*

Currently, we are expanding the system in different directions including: multiple-entities tracking, automatic generation of the passive radio map, location clustering, optimizing the APs and MPs positions, and analyzing the effect of dynamic changes in the environment and different hardware.

REFERENCES

[1] P. Enge and P. Misra. Special issue on GPS: The global positioning system. *Proceedings of the IEEE*, pages 3–172, January 1999.
[2] N. B. Priyantha, A. Chakraborty, and H. Balakrishnan. The cricket location-support system. In *6th ACM MOBICOM*, Boston, MA, August 2000.
[3] R. Want, A. Hopper, V. Falco, and J. Gibbons. The active badge location system. *ACM Transactions on Information Systems*, Vol. 10(1):91–102, January 1992.
[4] P. Bahl and V. N. Padmanabhan. RADAR: An In-Building RF-based User Location and Tracking System. In *IEEE Infocom 2000*, volume 2, pages 775–784, March 2000.
[5] Moustafa Youssef, Matthew Mah, and Ashok Agrawala. Challenges: device-free passive localization for wireless environments. In *MobiCom '07: Proceedings of the 13th annual ACM international conference on Mobile computing and networking*, pages 222–229, New York, NY, USA, 2007. ACM.
[6] Joey Wilson and Neal Patwari. Radio tomographic imaging with wireless networks. *Tech Report*, Sep 2008.
[7] M. Youssef and A. Agrawala. Small-scale compensation for WLAN location determination systems. In *IEEE WCNC 2003*, March 2003.
[8] M. Youssef and A. Agrawala. The Horus WLAN Location Determination System. In *Third International Conference on Mobile Systems, Applications, and Services (MobiSys 2005)*, June 2005.
[9] M. Youssef, A. Agrawala, and A. U. Shankar. WLAN location determination via clustering and probability distributions. In *IEEE PerCom 2003*, March 2003.
[10] M. Youssef and A. Agrawla. The horus location determination system. *ACM Wireless Networks (WINET) Journal*, 2007.
[11] M. Youssef and A. Agrawala. On the Optimality of WLAN Location Determination Systems. In *Communication Networks and Distributed Systems Modeling and Simulation Conference*, January 2004.
[12] P. Bahl, V. N. Padmanabhan, and A. Balachandran. Enhancements to the RADAR user location and tracking system. Technical Report MSR-TR-00-12, Microsoft Research, February 2000.
[13] Moustafa Seifeldin and Moustafa Youssef. Nuzzer: A large-scale device-free passive localization system for wireless environments. Technical Report Wireless Intelligent Networks Center (WINC) Technical Report WINC-TR2009-1007, 2009.
[14] IEEE Computer Society LAN MAN Standards Committee. Wireless LAN Medium Access Control (MAC) and Physical Layer (PHY) Specifications. In *IEEE Std 802.11-1999*, 1999.
[15] J. Krumm et al. Multi-camera multi-person tracking for easy living. In *3rd IEEE Int'l Workshop on Visual Surveillance*, pages 3–10, Piscataway, NJ, 2000.
[16] R. J. Orr and G. D. Abowd. The smart floor: A mechanism for natural user identification and tracking. In *Conference on Human Factors in Computing Systems (CHI 2000)*, pages 1–6, The Hague, Netherlands, April 2000.
[17] Y. Yang and A. E. Fathy. See-through-wall imaging using ultra-wideband short-pulse radar system. *Proc. IEEE Antennas Propag. Soc. Int. Symp. Dig.*, Vol. 3B:334–337, Jul. 2005.
[18] P. van Dorp and F. C. A. Groen. Human walking estimation with radar. *Proc. Inst. Electr. Eng.Radar, Sonar Navig.*, Vol. 150, no. 5:356 – 365, Oct. 2003.
[19] A. R. Hunt. A wideband imaging radar for through-the-wall surveillance. *Proc. SPIESensors, and Command, Control, Communications, and Intelligence (C3I) Technologies*, Vol. 5403:590 – 596, Sep. 2004.
[20] C. P. Lai and R. M. Narayanan. Through-wall imaging and characterization of human activity using ultrawideband (uwb) random noise radar. *Proc. SPIESensors and C3I Technologies for Homeland Security and Homeland Defense*, Vol. 5778:186 – 195, May 2005.
[21] A. Lin and H. Ling. Doppler and direction-of-arrival (ddoa) radar for multiple-mover sensing. *IEEE Trans. Aerosp. Electron. Syst.*, Vol. 43, no. 4:1496 – 1509, Oct. 2007.
[22] J. L. Geisheimer, E. F. Greneker, and W. S. Marshall. High-resolution doppler model of the human gait. *Proc. SPIERadar Sensor Technology and Data Visualization*, Vol. 4744:8 –18, Jul. 2002.
[23] S. S. Ram, Y. Li, A. Lin, and H. Ling. Human tracking using doppler processing and spatial beamforming. *IEEE 2007 Radar Conference*, 2007.
[24] A. M. Haimovich, R. S. Blum, and L. J. Cimini. Mimo radar with widely separated antennas. *IEEE Signal Processing Magazine*, pages 116–129, January 2008.